# A New Generalization of Edge Overlap to Weighted Networks

Ali Choumane

Lebanese University - Faculty of Sciences, LaRIFA Lab, Nabatieh – Lebanon

*ABSTRACT*

*Finding the strength of an edge in a network has always been a big demand. In the context of social networks, it allows to estimate the relationship strength between users. The best-known method to compute edge strength is the Neighbourhood Overlap. It computes the ratio of common neighbours to all neighbours of an edge terminal nodes. This method has been initially proposed for unweighted networks and later extended for weighted ones. These two versions of the method are not mathematically equivalent: In fact, an unweighted network is commonly considered as weighted with all edge weights equal to one. Using both existent versions of Neighbourhood Overlap on such network produce completely different values. In this paper, we tackle this problem and propose a new generalization for Neighbourhood Overlap that works equally for unweighted and weighted networks. Experiment performed on networks with various parameters showed similar performance of our measure to the existing measures.*

## 1. INTRODUCTION

A social network is a set of interconnected individuals, groups or organizations. For computational purposes, it is naturally represented by graph where a set of nodes are connected by edges. There are many analysis approaches that could be done to infer hidden information from the topology of the network only, without knowing the details of interaction between the nodes. One of the important analysis measures is the *Neighbourhood Overlap* (NO) [1]. NO measure assigns a weight to each edge in the network representing its strength. The strength and weakness in this context have special definitions [2, 3]. A strong edge is an edge in which its terminals share relatively high number of their neighbours. A weak edge is a local bridge in which its removal would make the shortest path between its terminals higher than 2, in other terms, they don't share any neighbour. NO gives a value for strength by calculating the ratio of common neighbours to all neighbours of both nodes, excluding themselves. Another approach is to weight the edge by the number of hops of the shortest path between its terminals if the edge were removed. This approach is computationally expensive, and it weights the edge according to its weakness not its strength, in other words, two nodes having 100 neighbours in common and two nodes with one common neighbour will have the same weight. NO proved to be useful in many applications. One of the most important is community detection [4, 1, 5, 6] where nodes are partitioned into subsets, each subset has a relatively larger internal edge density among its members than with the nodes of different subsets.

The early proposed NO measure works for unweighted networks only. According to our knowledge, only one extension to weighted networks has been proposed in [7]. As will be International Journal of Artificial Intelligence and Applications (IJAIA), Vol.11, No.1, January 2020International Journal of Artificial Intelligence and Applications (IJAIA), Vol.11, No.1, January 2020 explained in sections 2 and 3, this measure is not mathematically equivalent to the original measure if applied to unweighted network, i.e. if we consider edge weights to be all 1s or if an unweighted network is handled as weighted with weights being all 1s. In this case, the existing extension produces different values than the original NO. In this paper we propose a new measure that solves this point. It works equally for unweighted and weighted networks while having similar performance in complex networks compared to the existing measures.

The article is structured as follows. In section 2, we present a literature review related to edge strength computation in complex networks with focus on Neighbourhood Overlap and its applications. Section 3 presents the proposed measure and its mathematical characteristics. In section 4 we present an evaluation study that confirms the performance of our measure. Conclusion and future work are reported in section 5.

## 2. RELATED WORK

The neighbourhood overlap of an edge (i,j) [1] is defined as the number of nodes who are common neighbours of both i and j divided by the number of nodes who are neighbours of at least one of the two nodes i or j (excluding each other). It is equal to:

$$O_{ij} = \frac{c_{ij}}{k_i + k_j - 2 - c_{ij}} \tag{1}$$

where $k_i$ and $k_j$ are, respectively, the degrees of nodes i and j, and $c_{ij}$ is the number of common neighbours of both i and j. The factor of 2 in the denominator is to account for excluding vertices i and j who are neighbours of each other. Low values of this measure indicate weak ties between the corresponding nodes.

In [2], the authors defined strong and weak ties, and based on them, they used the formula above to quantify the strength of an edge by a real number instead of just being weak or strong. In [8], the authors confirmed the results by using a phone call data that allowed them to weight an edge by the number of minutes spent on a call between two nodes in the network.

They realized that the curve of neighbourhood overlap as function of their percentile in the sorted order of all edges is almost linear, which means that this measure could be actually used to infer tie strength. In [9] authors applied the notion of strong and weak edges in social networks, actually, they redefined the weak ties in a different way that suits better an online social network like Facebook.

NO measure was used in various community detection algorithms, in which some of them depends heavily on it. The algorithm proposed in [4] depends on the principle that low NO-weighted edges represent local bridges between edges. It removes edges in increasing order of their NO weight, thus removing the local bridges one after another, then finding the case that gives highest quality function value (called modularity) [10]. Another algorithm based on NO was introduced in [5], which uses the notion of weak ties to sort nodes into communities.

92

The existing NO measure (equation 1) computes the strength of an edge no matter the weight associated to it in the input graph. In weighted graphs, the weight of an edge between two nodes vehicles information about the strength of their relationship. Hence, computing NO in weighted graphs should involve both the weight of the edge along with the proportion of common neighbours of its terminal nodes. According to our knowledge, the only research that extended the concept of NO to weighted networks is the work proposed in [7]:

$$O_{ij}^w = \frac{\sum_{k \in N_i \cap N_j}(w_{ik}+w_{jk})}{S_i+S_j-2w_{ij}} \quad (2)$$

Where $N_i$ ($N_j$) is the set of neighbours of node i (j), $w_{ik}$ denotes the weight associated with the edge between nodes i and k, and $S_i$ ($S_j$) denotes the strength of node i (j), where $S_i = \sum_{k \in N_i} w_{ik}$. The authors of this measure showed its performance in estimating edge strengths. The main problem is that this measure is not equivalent to the original measure (equation 1) if all edge weights are 1s, i.e. similar to unweighted network. Consider the following unweighted graph where we consider edge weights to be all 1s.

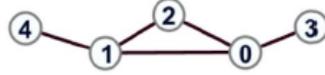

The NO of edge [1,2] is computed as follows:

$$O_{12}^w = \frac{w_{10} + w_{20}}{S_1 + S_2 - 2w_{12}} = \frac{1+1}{3+2-2} = \frac{2}{3}$$

$$\text{where } O_{12} = \frac{1}{3+2-2-1} = \frac{1}{2}$$

In this article we propose a new measure that solves this problem mathematically while maintaining the performance of estimating edge strengths.

## 3. PROPOSED MEASURE

Our proposed measure (called NO) is calculated using the following equation:

$$NO_{ij}^w = \frac{\sum_{k \in N_i \cap N_j} \min(w_{ik}, w_{jk})}{\sum_{k \in N_i \cup N_j - \{i,j\}} \max(w_{ik}, w_{jk})} \quad (3)$$

Where $N_i \cap N_j$ is the set of common neighbors of nodes $i$ and $j$, and $N_i \cup N_j - \{i,j\}$ is the union of the set of neighbors of both nodes, without i and j that are neighbours to each other.

This formula was meant to be a generalization to the neighbourhood overlap measure $O_{ij}$ on unweighted graphs (cf. equation 1). To achieve the generalization, we represent the network as sets of nodes, each set contains the neighbors of a node, the original neighbourhood overlap $O_{ij}$ between nodes $i$ and $j$ can then be written as the size of intersection over the size of union of neighbors of both nodes (excluding $i\ and\ j$ themselves):

$$\frac{|N_i \cap N_j|}{|(N_i-\{j\}) \cup (N_j-\{j\})|} \quad (4)$$

Let us define a partial belonging set (a set that its elements partially belong to it) as a mapping from the set of all nodes in the graph to the set of positive real numbers: $M: G \rightarrow R^+$ that maps each node to its membership in the set. Nodes outside the set are mapped to zero.

We define the size of a mapping to be $\sum_{x \in G} M(x)$, in our case it is the summation of all memberships of any node x in the set. In an unweighted network, where all nodes have memberships 1, this is equivalent to the cardinal of the set. In the case of a weighted network, we represent the set of neighbours $N_i$ of a node $i$ as a mapping $M_{Ni}$ defined as the following: $M_{N_i}(x) = w_{ix}$, thus, making the membership of the node $x$ to the set $N_i$ as the weight of the edge connecting it to $i$.

Now the problem of generalization has transformed into a problem of defining the intersection and union operators between two partial belonging sets, then it will be straight forward to replace the size of intersection between the neighbours of two nodes in equation 4 to the size of intersection between the partial belonging sets of the same two nodes, the same for union. In regular sets, the union between sets A and B is the set of elements which are in A, in B, or in both. In fact, it is the smallest set that is bigger than both. In the context of sets with partial belongings, the mapping corresponding for the union of A and B is the maximum of the mappings corresponding to A and B, because it is the only mapping that matches being the smallest mapping that is bigger than both:

$$\forall x \in (A \cup B), M_{A \cup B}(x) = \max(M_A(x), M_B(x)) \quad (5)$$

The intersection between sets A and B is the set that contains all the elements that are in both sets. It is the largest set that is smaller than both. In partial belonging sets, it is the minimum of the two mappings as it is the biggest mapping smallest than both, which is:

$$\forall x \in (A \cap B), M_{A \cap B}(x) = \min(M_A(x), M_B(x)) \quad (6)$$

International Journal of Artificial Intelligence and Applications (IJAIA), Vol.11, No.1, January 2020Accordingly, the neighbourhood overlap for a weighted edge (i,j) is the fraction of the size of the mapping $M_{N_i \cap N_j}$ to the size of the mapping $M_{N_i \cup N_j}$ that leads to the equation (3). The proposed measure is a generalization of the original neighbourhood overlap (equation 1 and 4), i.e. it gives same results in the case of unweighted graph (supposing the weight of each edge is 1). In fact, when the weights of all edges are ones, the minimum of weights of the two edges to each common neighbour will be also 1, thus the numerator will be equal to the count of common neighbours. Likewise, the denominator will be equal to the number of all neighbours without i and j themselves. In this case it gives the proportion of common neighbours to all neighbours similarly to the original measure.

## 4. EVALUATION

In order to evaluate our measure, we compared it to the reference measure $O_{ij}^w$, the only existing extension on weighted networks (cf. section 2). We compared the values obtained by our measure $NO_{ij}^w$ to $O_{ij}^w$ using the well-known Lancichinetti-Fortunato-Radicchi (LFR) benchmark. LFR benchmark is a reference algorithm that generates benchmark synthetic networks that resemble real-world networks [11]. This method allows to generate weighted networks and the underlying community structure that satisfy the user's parameters. Some of the parameters specify properties of communities in the network: n (number of nodes), $c_{min}$ and $c_{max}$ (minimum and maximum community size). The other parameters specify properties of the generated network: k (average degree), $k_{max}$ (maximum degree), $\mu_t$ (mixing parameter for the topology: each node shares a fraction $\mu_t$ of its edges with nodes in other communities), and $\mu_w$ (mixing parameter for the weights: each node shares a fraction $\mu_w$ of its total edge weights with nodes in other communities). We consider it is important to evaluate NO on such networks with well-known community structure as well as NO helps identifying communities as showed in section 2. In fact, it has been shown that NO is a significant feature that allows to partition the network into communities. NO is related to the network size and node degrees. Moreover, the distribution of the edge weights and the topology of the network should have high influence on the obtained values. Figures 1 and 2 show the results obtained by our measure and the reference measure $O_{ij}^w$ while varying the LFR parameters n, k, $\mu_t$, $\mu_w$. Each point is the plots is averaged over 50 networks. The main conclusion across all plots in both figures is that there is a linear relation between our measure and $O_{ij}^w$. Our measure is simply a linear translation of $O_{ij}^w$ and hence it can replace it in all circumstances. Moreover, our measure shows a consistent profile for all network sizes and mixing parameters where $O_{ij}^w$ seems to be sensitive for high mixing parameters ($\mu_t$ and $\mu_w$) in which it is hard to partition the network into communities. The obtained results confirm the validity of our measure in estimating the neighbourhood overlap of edges in weighted networks.

95

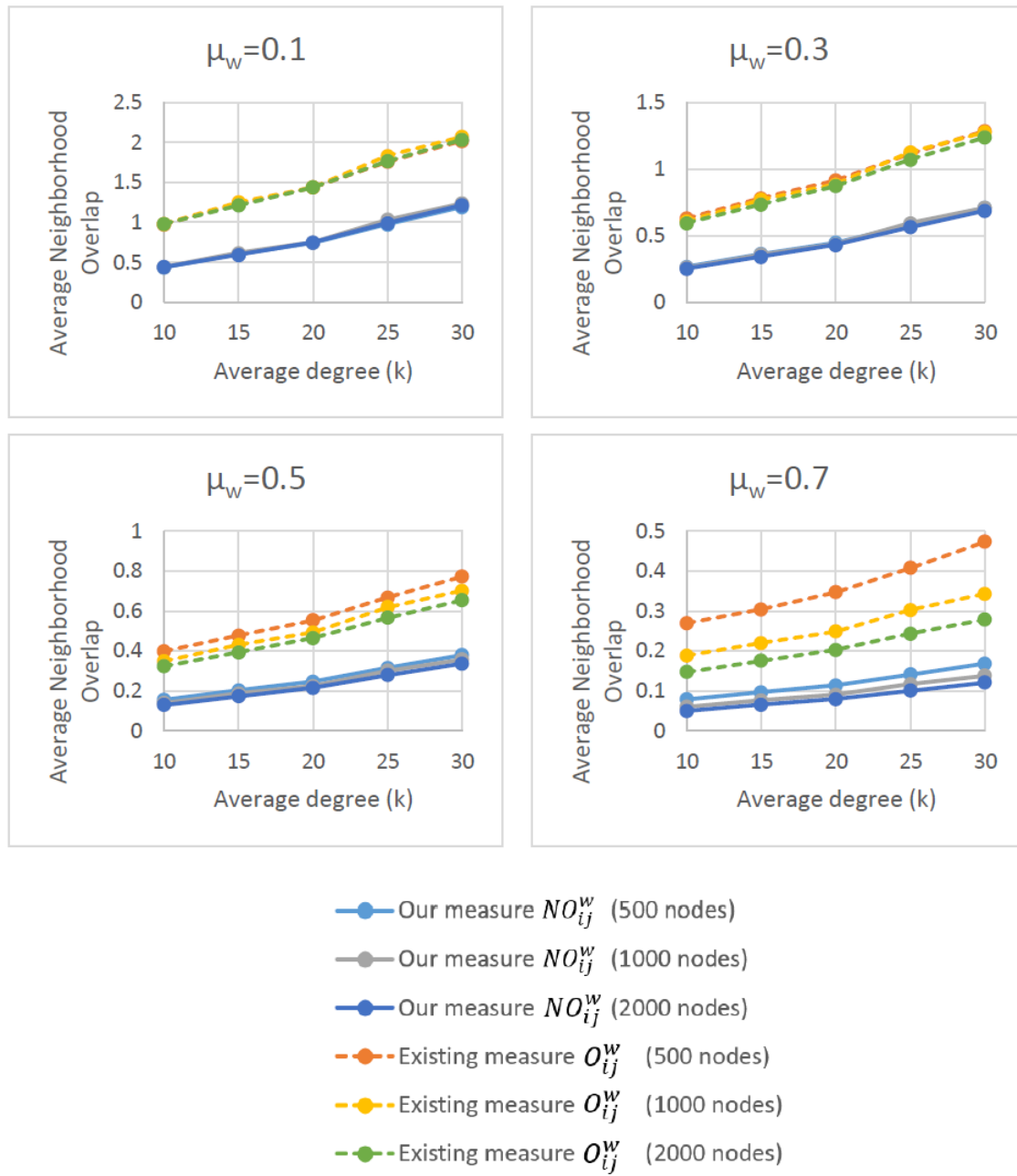

Figure. 1: The effects of network size, average degree and mixing parameter for the weights µw on LFR weighted networks. Plots show average neighbourhood overlap of the network according to our method and the method proposed in [7]. All results are from networks with mixing parameter for the topology = 0.3 and maximum degree = 50. All results are averaged over 50 networks with each set of parameters.

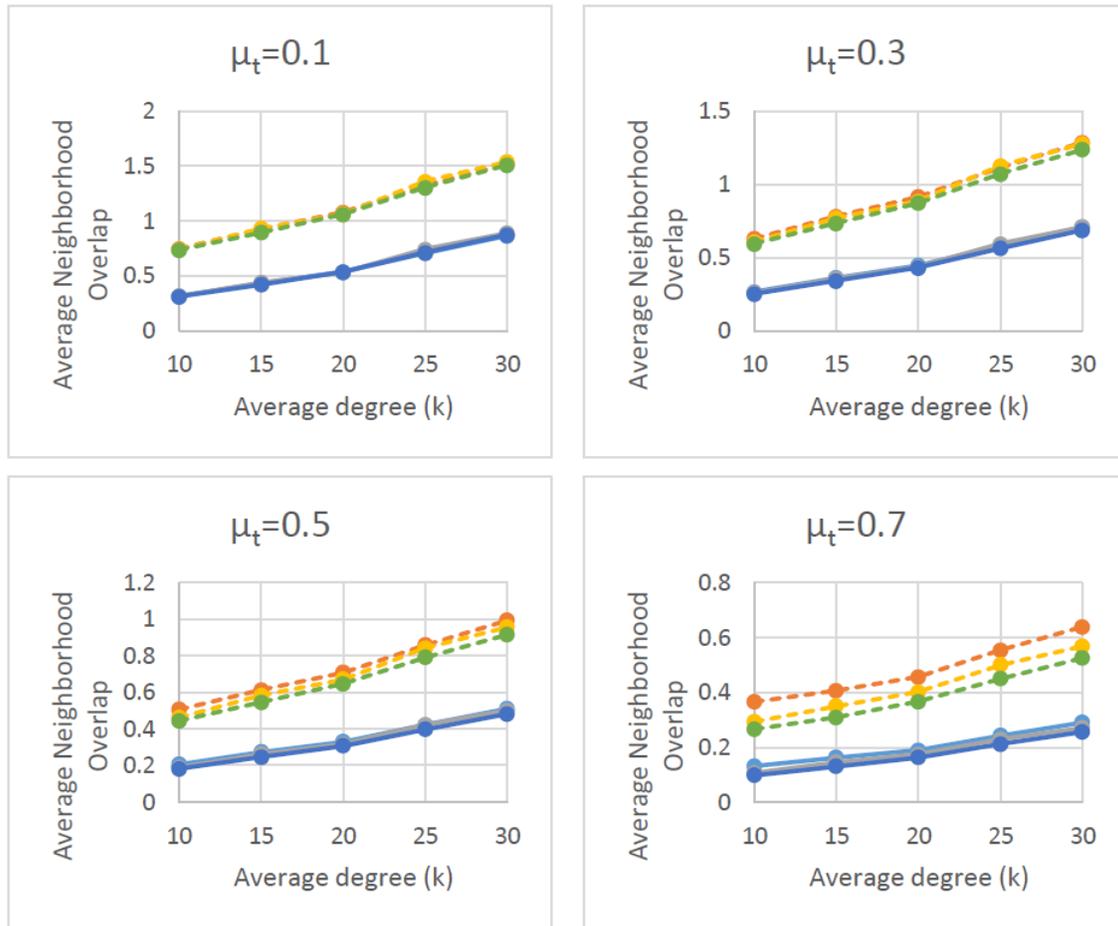

Figure. 2: The effects of network size, average degree and mixing parameter for the topology μt on LFR weighted networks. Plots show average neighbourhood overlap of the network according to our method and the method proposed in [7]. All results are from networks with mixing parameter for the weights = 0.3 and maximum degree = 50. All results are averaged over 50 networks with each set of parameters.

## 5. CONCLUSION AND FUTURE WORK

Neighbourhood Overlap measures edge strength in complex networks. In this article we proposed a new generalization of this measure in the context of weighted networks. In contrast to the existing measure $O_{ij}^w$, our measure $NO_{ij}^w$ can be applied to both weighted and unweighted networks as it is mathematically equivalent to original Neighbourhood Overlap if the edge weights are all 1s. Large experiment on the well-known LFR benchmark proved the validity of our measure. It has similar performance compared to the existing measure. Furthermore, our measure is stable across all network parameters while $O_{ij}^w$ is influenced by the topology and weights distribution of the network. As part of future work, we plan to validate our method on weighted directed networks and to use it in existing community detection algorithm.